\documentclass[a4paper]{article}

\usepackage{INTERSPEECH2022}
\usepackage{multirow,makecell,bm,cite,color,textcomp}

\title{MIMO-DoAnet: Multi-channel Input and Multiple Outputs DoA Network \\ with Unknown Number of Sound Sources}
\name{Haoran Yin$^{1,2}$, Meng Ge$^{1}$, Yanjie Fu$^{1}$, Gaoyan Zhang$^{1,*}$, Longbiao Wang$^{1,*}$,\\ Lei Zhang$^{2}$, Lin Qiu$^{2}$, Jianwu Dang$^{1,3}$
\thanks{* denotes the corresponding author.}}
\address{
  $^{1}$Tianjin Key Laboratory of Cognitive Computing and Application,\\
  College of Intelligence and Computing, Tianjin University, Tianjin, China\\
  $^{2}$ICT Products \& Solutions, Huawei, Dongguan, China\\
  $^{3}$Japan Advanced Institute of Science and Technology, Ishikawa, Japan}
\email{$\{$haoran\_yin, zhanggaoyan, longbiao\_wang$\}$@tju.edu.cn}

\begin{document}

\maketitle
\begin{abstract}
Recent neural network based Direction of Arrival (DoA) estimation algorithms have performed well on unknown number of sound sources scenarios. These algorithms are usually achieved by mapping the multi-channel audio input to the single output (i.e. overall spatial pseudo-spectrum (SPS) of all sources), that is called MISO. However, such MISO algorithms strongly depend on empirical threshold setting and the angle assumption that the angles between the sound sources are greater than a fixed angle.
To address these limitations, we propose a novel multi-channel input and multiple outputs DoA network called MIMO-DoAnet. Unlike the general MISO algorithms, MIMO-DoAnet predicts the SPS coding of each sound source with the help of the informative spatial covariance matrix. By doing so, the threshold task of detecting the number of sound sources becomes an easier task of detecting whether there is a sound source in each output, and the serious interaction between sound sources disappears during inference stage. Experimental results show that MIMO-DoAnet achieves relative 18.6\% and absolute 13.3\%, relative 34.4\% and absolute 20.2\% F1 score improvement compared with the MISO baseline system in 3, 4 sources scenes. The results also demonstrate MIMO-DoAnet alleviates the threshold setting problem and solves the angle assumption problem effectively.

\end{abstract}

\vspace{-5px}
\section{Introduction}

Direction of Arrival (DoA) estimation is an intelligent technology of estimating the direction of sound to the microphone array, which can provide the informative spatial property for many downstream tasks, such as speaker verification\cite{he2021speaker}, speech recognition\cite{subramanian2022recognition,kim2015recognition} and speech separation \cite{ge2022spex,zhang2021ADL,xu2021GRNN,li2021MIMO,chazan2019multi}.
With the further development of DoA estimation algorithms, more researchers aim to improve the robustness and adaptability of the algorithms in real scenes\cite{bohlender2021exploiting,gelderblom2021synthetic,grumiaux2021EUSIPCO,sundar2020raw,takeda2016discriminative}.

\begin{figure}[t]
		\begin{minipage}[h]{0.45\linewidth}
			\centering
			\centerline{\includegraphics[width=1\linewidth]{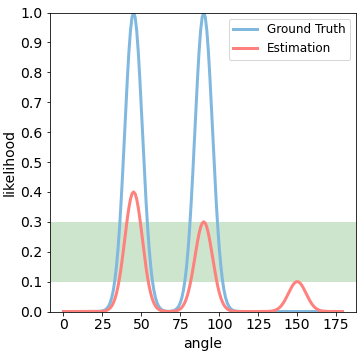}}
			\vspace{-5px}
			\centerline{(a) Threshold setting problem}\medskip
		\end{minipage}
		\hfill
		\begin{minipage}[h]{0.45\linewidth}
			\centering
			\centerline{\includegraphics[width=1\linewidth]{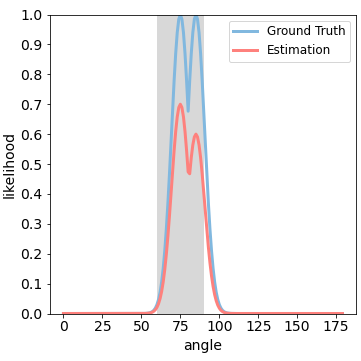}}
			\vspace{-5px}
			\centerline{(b) Angle assumption problem}\medskip
		\end{minipage}
		\vspace{-10px}
		\caption{Two examples of threshold setting problem and angle assumption problem, \textcolor[rgb]{0.5058,0.7215,0.9137}{blue} and \textcolor[rgb]{0.9960,0.5058,0.4901}{red} line denote \textcolor[rgb]{0.5058,0.7215,0.9137}{ground truth} and \textcolor[rgb]{0.9960,0.5058,0.4901}{estimated} spatial pseudo-spectrum (SPS), respectively.}
		\vspace{-25px}
		\label{fig:mix_examples}
\end{figure}

DoA estimation for unknown multiple sound sources is one of the challenges of DoA estimation algorithms in real scenes. Benefit from the strong non-linear mapping ability of neural networks, neural network based DoA estimation algorithms outperformed traditional algorithms \cite{daniel2020time,xu2012high, tervo2009direction,dmochowski2007generalized}. The neural network based DoA estimation algorithms \cite{le2021data,kapka2019sound} usually apply a threshold to count the number of sound sources in the inference stage and they achieve great performance on unknown multiple sound sources scenarios.
These algorithms widely use spatial pseudo-spectrum (SPS) as the output feature and map the multi-channel audio input to the single output (i.e. overall SPS of all sources), called MISO DoA network \cite{he2021DOA,nguyen2020robust,he2018joint,adavanne2018direction,he2018deep}. However, there are two problems limit the performance of the algorithms. 
The first problem is the varied threshold has a great impact on the estimation of the number of sound sources, as shown in Figure \ref{fig:mix_examples}(a), the DoA algorithms estimate the number of sound sources correct only when the threshold is set in the green range, but the fixed empirical threshold could not handle the changeable SPS conditions. The second problem is these algorithms always rely on an assumption that the angle between every two sound sources is greater than the half beam width of SPS. When the DoA estimation algorithms deal with two close sound sources in the inference stage, they find the highest peak of SPS and calculate the first estimation angle, then they set the likelihood of angles within the SPS beam of the first angle to zero, as shown in the gray range of the Figure \ref{fig:mix_examples}(b), result in the second angle is ignored and a wrong estimation of the source number.

To alleviate the problems, we propose a multi-channel input and multiple outputs DoA network called MIMO-DoAnet.
MIMO-DoAnet separates the mixture into single speech using the complex-valued ratio filter (cRF) \cite{mack2019CRF} in the cRF estimator, then calculates covariance matrices of each single speech and predicts the multiple DoA of each sound source through the SPS estimators. In this way, the task of threshold is changed from detecting the number of sound sources to detect whether there is a sound source in each output, and there is no interference between sound sources. With the help of covariance matrices and multiple outputs network, MIMO-DoAnet alleviates the threshold setting problem and overcomes the angle assumption problem effectively.
The main advantages of MIMO-DoAnet are summarized in the following 3 points.

\begin{figure*}[t]
		\begin{minipage}[h]{1\linewidth}
		\vspace{-8px}
			\centering
			\centerline{\includegraphics[width=0.75\linewidth]{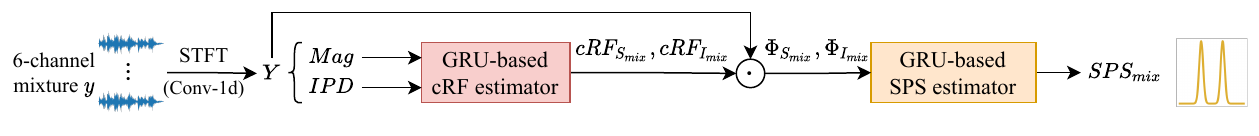}}
			\vspace{-8px}
			\centerline{(a) The framework of MISO DoA baseline}\medskip
		\end{minipage}
		\begin{minipage}[h]{1\linewidth}
			\vspace{-8px}
			\centering
			\centerline{\includegraphics[width=0.75\linewidth]{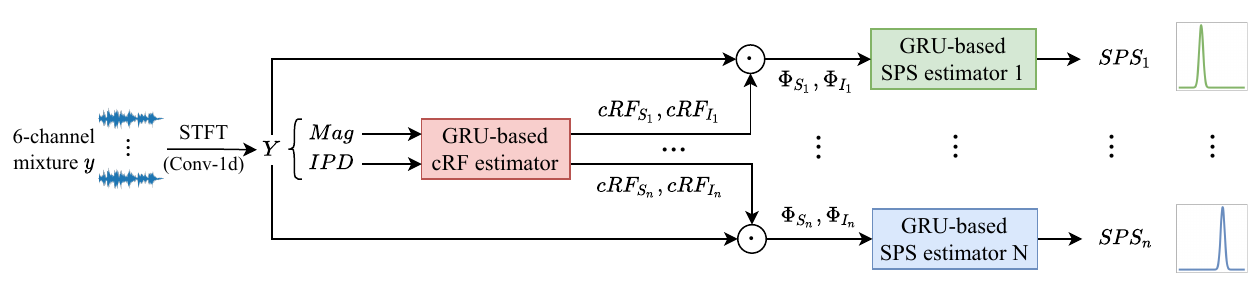}}
			\vspace{-11px}
			\centerline{(b) The framework of MIMO-DoAnet}\medskip
		\end{minipage}
		\vspace{-11px}
		\caption{The frameworks include GRU-based cRF estimator and GRU-based SPS estimator, $y$ denotes the 6-channel input waveform, $Y$ represents the features obtained by $y$ through a fixed STFT encoder\cite{gu2020multi}. $\text{Mag}$ and $\text{IPD}$ denote the magnitude spectrogram of the first channel and the interaural phase difference. $N$ denotes the maximum number of sound sources. The complex-valued ratio filter of speech $\text{cRF}_{S}$ and interference $\text{cRF}_{N}$ are estimated by GRU-based cRF estimator, and the covariance matrix $\Phi$ is calculated by the $\text{cRF}$ and $Y$ (as shown in Eq.(\ref{eq:speech}) and (\ref{eq:covariance})). In the GRU-based SPS estimator, SPS is predicted from the concatenation of $\Phi_{S}$ and $\Phi_{I}$.}
		\vspace{-15px}
		\label{fig:sturcture}
\end{figure*}

First, MIMO-DoAnet has a strong adaptability to the DoA estimation of the unknown number of sound sources. Different from MISO DoA algorithms that suffer from drastic performance degradation with more sound sources, MIMO-DoAnet gets outstanding performance in the case of 2 sources, 3 sources, and even 4 sources.
Second, MIMO-DoAnet reduces the difficulty in threshold setting. The threshold task of detecting whether there is a sound source in each output is much easier so that MIMO-DoAnet easily finds a suitable threshold setting.
Third, MIMO-DoAnet has a strong ability to estimate close sound sources. In the small included angle test set, MIMO-DoAnet performs much better than MISO DoA algorithms.
By the way, MIMO-DoAnet has low-latency benefited from the 32 ms input feature and gated recurrent unit (GRU)\cite{cho2014learning} network.



\vspace{-5px}
\section{MISO DoA network}
\vspace{-2px}

Since the input feature and network structure of DoA networks are very different, we design a MISO DoA network as the baseline to verify the effectiveness of MIMO-DoAnet in solving the threshold setting problem and angle assumption problem. The input feature, the cRF estimator, and the SPS estimator of MISO DoA network are almost the same as that of MIMO-DoAnet, and the output feature and inference process are the same as existing neural network based DoA estimation algorithms.

\vspace{-5px}
\subsection{Input and output features}
\vspace{-2px}
We concatenate the magnitude spectrogram of the first channel mixture speech and the interaural phase difference (IPD) as the input feature of MISO DoA network.
We estimate the DoA for sound sources from 0$^{\circ}$ to 180$^{\circ}$ of the linear microphone array. Existing methods generally consider a sound source to be dominant within a range of 30$^{\circ}$ to the left and right of it, which we call the beam width. To maintain the integrity of the SPS beam at the edge area, we extend 15$^{\circ}$ to the left of 0$^{\circ}$ and to the right of 180$^{\circ}$, so we apply a 210-D SPS. Correspondingly, we represent an angle with the SPS of the angle+15$^{\circ}$. The SPS is encoded as the following likelihood $p(\theta_i)$, where $i$ ranges from 0 to 210, $\theta_i$ denotes the $i$-th index of the SPS:
\vspace{-5px}
\begin{equation}
p(\theta_i)=\begin{cases}\max _{\theta^{\prime} \in x}\left\{\exp \left(-\frac{d\left(\theta_i,\theta^{\prime}+15\right)^{2}}{\sigma^{2}}\right)\right\} & \text { if }|x|>0 \\ 0 & \text { otherwise }\end{cases}
\label{}
\end{equation}
where $x$ denotes the set of the ground truth sound sources, $|x|$ denotes the number of sources, $\theta^{\prime}$ is the ground truth angle of each source, $\sigma$ is a pre-defined parameter for the beam width and $d(\cdot,\cdot)$ denotes the angular distance.

\subsection{Network structure}

As shown in Figure \ref{fig:sturcture}(a), the MISO DoA network consists of a GRU-based cRF estimator and a GRU-based SPS estimator.
The complex-valued ratio filter (cRF) is an extended version of complex-valued ratio mask (cRM)\cite{williamson2015CRM} to calculate the target speech and interference covariance matrices. The GRU-based cRF estimator predicts the cRF of speech $\text{cRF}_{S}$ and interference $\text{cRF}_{I}$, the estimated speech $\tilde{S}(t,f)$ is calculated as:
\vspace{-6px}
\begin{equation}
\tilde{S}(t,f) = \sum_{\tau_{1}=-L}^{\tau_{1}=L}\sum_{\tau_{2}=-L}^{\tau_{2}=L}\text{cRF}_{S}(t+\tau_{1},f+\tau_{2})*Y(t+\tau_{1},f+\tau_{2})
\vspace{-3px}
\label{eq:speech}
\end{equation}
where $t$ and $f$ denote the time frame index and the frequency bin index respectively, $\tau_1$ and $\tau_2$ denote the considered regions in the time axis and frequency axis range, $Y$ denotes the complex spectrums of the multi-channel mixture, $L$ is the neighboring context size. The corresponding estimated interference $\hat{I}(t,f)$ is calculated with $\text{cRF}_{I}$ in the same way. The calculation of frame-wise speech covariance is as follows:
\vspace{-3px}
\begin{equation}
\Phi_{S}(t,f)=\frac{\tilde{S}(t,f)\tilde{S}^{H}(t,f)}{\sum_{t=1}^{T}\text{cRM}_{S}^{H}(t,f)\text{cRM}_{S}(t,f)}
\vspace{-3px}
\label{eq:covariance}
\end{equation}
where $T$ denotes the number of time frames, $\text{cRM}_{S}(t,f)$ denotes the center mask of $\text{cRF}_{S}(t,f)$, and $H$ denotes the conjugate transpose. Due to the MISO DoA network estimates all the DoA in one SPS, so the cRF and covariance matrices of speech and interference are represented by $\text{cRF}_{S_{mix}}$, $\text{cRF}_{I_{mix}}$ and $\Phi_{S_{mix}}$, $\Phi_{I_{mix}}$, respectively.
Then the $\Phi_{S_{mix}}$ and $\Phi_{I_{mix}}$ are concatenated as the input of the GRU-based SPS estimator. The loss function of MISO DoA network $L_{\text{MISO}}$ is as follows:
\vspace{-3px}
\begin{equation}
L_{\text{MISO}} = \sum_{t=1}^{T}(\widetilde{\text{SPS}}_\text{mix}(t)-\text{SPS}_\text{mix}(t))^2
\vspace{-8px}
\end{equation}
where $\widetilde{\text{SPS}}_\text{mix}(t)$ denotes the frame-wise estimated SPS, $\text{SPS}_\text{mix}(t)$ denotes the frame-wise ground truth SPS.

\vspace{-3px}
\subsection{Inference process}

Like existing threshold-based DoA methods, the first step is setting a threshold. Then select the corresponding angles of all the SPS likelihood greater than the threshold as the candidate angle set. The next step is to find the angle corresponding to the highest peak of likelihood in the candidate angle set. Then put the angle into the DoA prediction set and set the likelihood of angles within the predicted angle $\pm$15$^{\circ}$ to zero. Repeat the above process until the candidate set is empty, then we subtract 15$^{\circ}$ from the angles in the prediction set to get the DoA results.

\section{MIMO DoA network}

The GRU-based cRF estimator of MIMO-DoAnet predicts multiple cRF for multiple sound sources, and the multiple covariance matrices are calculated as the inputs of the GRU-based SPS estimators to predict multiple SPS for each sound source.

\vspace{-3px}
\subsection{Angle sorting algorithm}

MIMO-DoAnet has multiple outputs, so the SPS of each sound source is encoded separately.
In the unknown number of sound sources scenes, we need to design an algorithm to supervise the MIMO-DoAnet to know which angle each branch should predict in each frame. So we propose an angle sorting algorithm, we first get the voice activity detection (vad) label of each single source speech through a vad algorithm, then we calculate the angles of non-silent sound sources in each frame. We set all-zero SPS as the learning target of silent sound sources and we assume the all-zero SPS is the smallest in the sorting stage. Finally, we sort the angles of sound sources from smallest to largest and map each angle to each branch in order.

\vspace{-3px}
\subsection{Network structure}

As shown in Figure \ref{fig:sturcture}(b), the MIMO-DoAnet has $N$ GRU-based SPS estimators and $N$ outputs to estimate each sound source, $N$ denotes the maximum number of sound sources. The GRU-based cRF estimator predicts $N$ sets of cRF, and $N$ sets of covariance matrices are calculated as the input of $N$ GRU-based SPS estimators separately. Each GRU-based SPS estimator of MIMO-DoAnet has the same structure as the GRU-based SPS estimator of MISO baseline. The loss function of each output of MIMO-DoAnet is calculated in the same way as $L_{\text{MISO}}$, and we use different symbols to represent the loss function:
\vspace{-5px}
\begin{equation}
L_{\text{MIMO}_i} = \sum_{t=1}^{T}(\widetilde{\text{SPS}}_{i}(t)-\text{SPS}_{i}(t))^2
\vspace{-5px}
\end{equation}
where $i$ denotes the $i$-th output of MIMO-DoAnet, the loss function of entire MIMO-DoAnet $L_{\text{MIMO}}$ is given as follows:
\vspace{-8px}
\begin{equation}
L_{\text{MIMO}} = \sum_{i=1}^{n}L_{\text{MIMO}_i}
\vspace{-5px}
\end{equation}
where $n$ denotes the maximum number of sound sources.

\vspace{-3px}
\subsection{Inference process}

We also set a threshold to help MIMO-DoAnet determine whether there is a sound source in each output. If the maximum likelihood of SPS is greater than the threshold, we put the highest peak corresponding angle into DoA estimation results, otherwise, we ignore the output.

\begin{table}[t]
\centering
\caption{The parameters setting of small, middle, large rooms.}
\vspace{-8px}
    \setlength{\tabcolsep}{1.2mm}{
        \begin{tabular}{c|c|c|c|c}
        \toprule
        \textbf{Size}& \textbf{Length}$(L)$ & \textbf{Width}$(W)$ & \textbf{Height}$(H)$ & \textbf{RT60} \\
        \midrule
        \textbf{Small} & \lbrack 4m, 6m\rbrack  & \multirow{3}{*}{\lbrack 3m, $L$\rbrack } &  \multirow{3}{*}{\lbrack 3m, 3.5m\rbrack } & \lbrack 0.2s, 0.5s\rbrack  \\
        \textbf{Middle} & \lbrack 6m, 10m\rbrack  & & & \lbrack 0.3s, 0.6s\rbrack  \\
        \textbf{Large} & \lbrack 10m, 15m\rbrack & & & \lbrack 0.4s, 0.7s\rbrack  \\
        \bottomrule
        \end{tabular}}
\vspace{-10px}
\label{tbl:room_params}
\end{table}

\section{Dataset and experimental setup}

\subsection{Dataset}
VCTK corpus\cite{yamagishi2019cstr} includes 48k Hz 2-channel speech data uttered by 110 English speakers, the length of data is concentrated in 3 seconds to 4 seconds, so we choose all the first channel of 4-second length speech data as original single-channel speech, and we downsample the sampling rate to 16k Hz. 

We simulate 6-channel speech data from original single-channel data using pyroomacoustics\cite{scheibler2018pyroomacoustics}, the spacing of the 6 linear microphones is 0.04 m, 0.04 m, 0.12 m, 0.04 m, 0.04 m. The parameters of simulated rooms are shown in Table \ref{tbl:room_params}.

We place the microphone array in the middle of the wall, at a distance of 0.5 m from the wall and 2 m from the ground. To make sound sources cover the area in rooms fully, we first set the angle of the source, then we leave 0.5 m between the sound source and the microphone array and between the sound source and the wall, divide the rest range into near, medium and far range, the distance between microphone array and the sound source is a random number in 3 types of range, so we simulate one single channel speech data at 3 distances simultaneously.

Then we generate 3 sets of training, validation set, and testing set from simulated rooms for 2, 3, and 4 sources respectively. We randomly select the angles and the near, medium, and far distance of sound sources, and keep the sound sources are at least 5$^{\circ}$ apart from each other. 
The training set contains 40,000 utterances (44.44 hours, 90 speakers), the validation set and the testing set contain 1,000 utterances (1.11 hours, 10 speakers) separately. Our simulated dataset can be found at https://github.com/TJU-haoran/VCTK-16k-simulated.git

\vspace{-5px}
\subsection{Experimental setup}

We implement our experiments by using Pytorch 1.8.0. The 512 frame size Hamming window is used with 50\% overlap. The $\sigma$ is set to 8 and the $L$ of cRF is set to 3. The network is trained using one warm-up epoch and using Adam optimizer\cite{kingma2014adam} with early stopping. The initial learning rate is set to 1e-4 and the gradient norm is clipped with max norm 3. The cRF estimator in MIMO-DoAnet includes a 256 units fully connected (FC) layer, a uni-directional 2 layers 500 units GRU with Relu activation function and $2N$ 4626 units FC layers for $\Phi_{S_1}...\Phi_{S_n}$ and $\Phi_{I_1}...\Phi_{I_n}$ with layer normalization \cite{ba2016layer}. The SPS estimator consists of a 300 units FC layer, a uni-directional 2 layers 300 units GRU, and a 210 units FC output layer.

\vspace{-6px}
\section{Results and discussions}
\vspace{-1px}

Since there are too many differences between MIMO-DoAnet and existing methods such as input feature and network type, we only conduct comparative experiments with the MISO baseline.
In this section, we analyze the overall performance, the effect of threshold setting, and the performance on small included angle test set. We evaluate the performance of MIMO-DoAnet and MISO baseline by using Recall, Precision, and F1 score\cite{1992F1Score}, we consider the error of DoA estimation less than 5$^{\circ}$ as correct.

\begin{table*}[t]
\centering
\small
\caption{Experiment results of MISO DoA baseline (MISO) and our proposed MIMO-DoAnet (MIMO) on test sets.}
\vspace{-5px}
    \setlength{\tabcolsep}{1.2mm}{
        \begin{tabular}{c c p{1cm}<{\centering}p{1.2cm}<{\centering}p{1.2cm}<{\centering}p{1cm}<{\centering}p{1.2cm}<{\centering}p{1.2cm}<{\centering}p{1cm}<{\centering}p{1.2cm}<{\centering}p{1.2cm}<{\centering}}
        \toprule
        \multirow{2}{*}{\textbf{SIR}} & \multirow{2}{*}{\textbf{Model}} & \multicolumn{3}{c}{\textbf{2 sources $(N=2)$}} &  \multicolumn{3}{c}{\textbf{3 sources $(N=3)$}} &  \multicolumn{3}{c}{\textbf{4 sources $(N=4)$}} \\\cmidrule(l){3-5}\cmidrule(l){6-8}\cmidrule(l){9-11}
        & & \textbf{Recall} & \textbf{Precision} & \textbf{F1 Score} & \textbf{Recall} & \textbf{Precision} & \textbf{F1 Score} & \textbf{Recall} & \textbf{Precision} & \textbf{F1 Score} \\
        \midrule

		\multirow{2}{*}{0} & MISO & 0.9050 & 0.9077 & 0.9064 & 0.8391 & 0.6243 & 0.7159 & 0.7796 & 0.4695 & 0.5861 \\
		& MIMO & \textbf{0.9059} & \textbf{0.9301} & \textbf{0.9179} & \textbf{0.8394} & \textbf{0.8592} & \textbf{0.8492} & \textbf{0.7810} & \textbf{0.7950} & \textbf{0.7880} \\
		\midrule

		\multirow{2}{*}{[-10, 10]} & MISO & 0.8871 & \textbf{0.9182} & \textbf{0.9024} & \textbf{0.8296} & 0.6138 & 0.7056 & 0.7445 & 0.4611 & 0.5695 \\
		& MIMO & \textbf{0.8886} & 0.9164 & 0.9023 & 0.8294 & \textbf{0.8436} & \textbf{0.8364} & \textbf{0.7641} & \textbf{0.7764} & \textbf{0.7702} \\ \bottomrule
        \end{tabular}}
\vspace{-10px}
\label{tbl:entire_results}
\end{table*}

\vspace{-6px}
\subsection{Overall performance}
\vspace{-1px}


The upper two rows of Table \ref{tbl:entire_results} present experimental results without adjustment for source-to-interferences ratio (SIR) \cite{vincent2006performance}, MIMO-DoAnet achieves relative 1.3\% and absolute 1.1\%, relative 18.6\% and absolute 13.3\%, relative 34.4\% and absolute 20.2\% F1 score improvement in 2, 3 and 4 sources scenarios. The sharp drop in precision leads to a decrease in the overall performance of the MISO baseline, which indicates the output SPS includes many wrong estimated angles in 3 and 4 sources scenarios. MIMO-DoAnet estimates one sound source in each output, so the outputs of MIMO-DoAnet are not allowed to contain more estimated angles. The lower two rows of Table \ref{tbl:entire_results} show the results with a random SIR $\in$ [-10, 10] of different sound sources, the performance of both MIMO-DoAnet and MISO baseline drops slightly, indicating that MIMO-DoAnet and MISO baseline could adapt the sound sources at different volumes with the help of spatial covariance matrices.


\begin{figure}[h]
		\begin{minipage}[h]{0.48\linewidth}
			\centering
			\centerline{\includegraphics[width=1\linewidth]{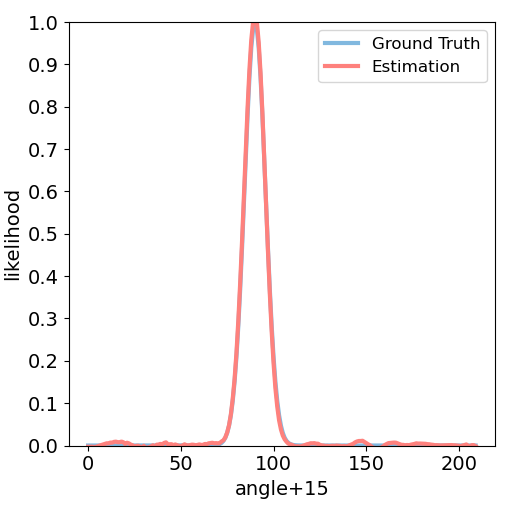}}
			\vspace{-3px}
			\centerline{(a) non-silent sound source}\medskip
		\end{minipage}
		\hfill
		\begin{minipage}[h]{0.48\linewidth}
			\centering
			\centerline{\includegraphics[width=1\linewidth]{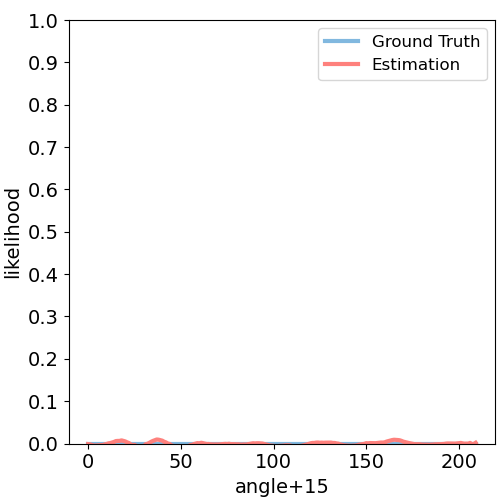}}
			\vspace{-3px}
			\centerline{(b) silent sound source}\medskip
		\end{minipage}
		\vspace{-10px}
		\caption{Two SPS outputs of MIMO-DoA network (2 sources).}
		\vspace{-5px}
		\label{fig:two_examples}
\end{figure}

\begin{table}[h]
\centering
\small
\caption{Experiment results of MISO-DoA baseline (MISO) and MIMO-DoA network (MIMO) on 2 sources (SIR=0) with different thresholds from 0.1 to 0.9, $\xi$ denotes the threshold.}
\vspace{-5px}
    \setlength{\tabcolsep}{1.2mm}{
    \resizebox{0.9\columnwidth}{23mm}{
        \begin{tabular}{c c c c c c c}
        \toprule
        \multirow{2}{*}{\textbf{$\xi$}} & \multicolumn{2}{c}{\textbf{Recall}} &  \multicolumn{2}{c}{\textbf{Precision}}  &  \multicolumn{2}{c}{\textbf{F1 Score}}  \\\cmidrule(l){2-3}\cmidrule(l){4-5}\cmidrule(l){6-7}
        & MISO & MIMO & MISO & MIMO & MISO & MIMO\\
        \midrule
        0.1&\textbf{0.9337}&\textbf{0.9059}&0.7579&0.9301&0.8367&\textbf{0.9179}\\
        0.2&0.9183&0.8993&0.8817&0.9313&0.8996&0.9151\\
        0.3&0.9050&0.8946&0.9077&0.9323&\textbf{0.9064}&0.9130\\
        0.4&0.8917&0.8900&0.9204&0.9332&0.9058&0.9111\\
        0.5&0.8779&0.8846&0.9295&0.9350&0.9030&0.9091\\
        0.6&0.8602&0.8777&0.9389&0.9377&0.8978&0.9067\\
        0.7&0.8342&0.8680&0.9496&0.9411&0.8882&0.9030\\
        0.8&0.7823&0.8530&0.9617&0.9462&0.8627&0.8972\\
        0.9&0.6428&0.8059&\textbf{0.9783}&\textbf{0.9583}&0.7759&0.8755\\
        \bottomrule
        \end{tabular}}}
\vspace{-10px}
\label{tbl:thresholds}
\end{table}

\subsection{The effect of threshold setting}

\begin{figure}[h]
\centering
	\includegraphics[width=0.6\linewidth]{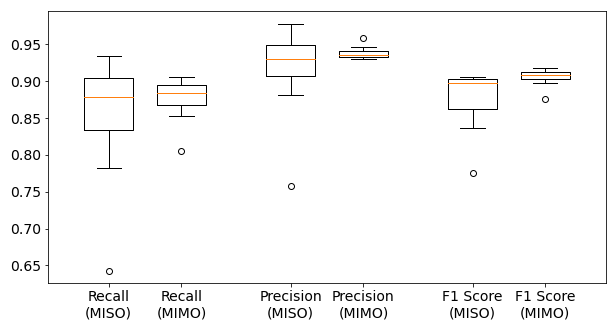}
	\vspace{-10px}
	\caption{The box plot of the experiment results in Table 3.}
	\label{fig:boxplot}
	\vspace{-6px}
\end{figure}

Due to the limited space, we only present the experimental results of MISO DoA baseline and MIMO-DoAnet on 2 sources test set (SIR=0) in this part. As shown in Figure \ref{fig:two_examples}, the estimated SPS is almost the same as ground truth SPS of both non-silent and silent sound sources, the obvious difference between output SPS of the non-silent and the silent sound source makes the threshold task of detecting whether there is a sound source much easier.
The experimental results with different thresholds from 0.1 to 0.9 are shown in Table \ref{tbl:thresholds}. The best F1 score performances of MIMO-DoAnet are implemented with the 0.1 thresholds in all of our experiments, but the thresholds for the best performances of MISO baseline fluctuate between 0.3 and 0.6.
The distribution of recall, precision, and F1 score of MIMO-DoAnet are more densely than that of MISO DoA baseline as shown in Figure \ref{fig:boxplot}, it shows that the impact of the threshold setting on the MIMO-DoAnet is much smaller than that of MISO DoA baseline, and MIMO-DoAnet always gets the best performance with a precise threshold of 0.1 instead of an empirical setting.

\begin{table}[h]
\centering
\small
\caption{Experiment results of MISO DoA baseline (MISO) and MIMO-DoAnet (MIMO) on small included angle test sets.}
\vspace{-8px}
    \setlength{\tabcolsep}{1.2mm}{
    \resizebox{0.8\columnwidth}{14mm}{
        \begin{tabular}{c c c c c}
        \toprule
        \textbf{Model} & \textbf{Source Number} &\textbf{Recall}& \textbf{Precision} & \textbf{F1 Score} \\
        \midrule
        \multirow{3}{*}{MISO} & 2 sources & 0.7039 & 0.8332 & 0.7632 \\
        & 3 sources & 0.7628 & 0.4935 & 0.5993 \\
        & 4 sources & 0.7712 & 0.4446 & 0.5640 \\
        \midrule
        \multirow{3}{*}{MIMO} & 2 sources & \textbf{0.8931} & \textbf{0.9224} & \textbf{0.9075} \\
        & 3 sources & \textbf{0.8146} & \textbf{0.8276} & \textbf{0.8211} \\
        & 4 sources & \textbf{0.7979} & \textbf{0.8110} & \textbf{0.8044} \\
		\bottomrule
        \end{tabular}}}
\vspace{-15px}
\label{tbl:small_test_set}
\end{table}

\vspace{-5px}
\subsection{The performance on small included angle test set}
\vspace{-2px}

We select 50 speech data with an included angle between two sources less than 15$^{\circ}$ from the test sets of 2, 3, and 4 sources. As shown in Table \ref{tbl:small_test_set}, MIMO-DoAnet achieves relative 18.9\% and absolute 14.4\%, relative 37\% and absolute 22.2\%, relative 42.6\% and absolute 24\% F1 score improvement in 2, 3 and 4 sources small included angle test set. Limited by the serious interaction between the two close sound sources during the inference stage, the MISO DoA baseline suffers a severe performance drop compared with the results on entire test sets (Table \ref{tbl:entire_results}), especially in 2 and 3 sources scenes. But the performance of MIMO-DoAnet is stable and even better, it shows MIMO-DoAnet overcomes the angle assumption problem.

\vspace{-5px}
\section{Conclusions and future work}

This paper proposes a novel multi-channel input and multiple outputs DoA network (MIMO-DoAnet) to address the limitations of threshold setting problem and overcome the angle assumption problem commonly in existing multi-channel input and single output (MISO) methods with the unknown number of sound sources. Benefitting from the combination of covariance matrices and multiple outputs for each sound source, the task of the threshold is much easier and the serious interaction between sound sources disappears. MIMO-DoAnet achieves relative 18.6\% and absolute 13.3\%, relative 34.4\% and absolute 20.2\% F1 score improvement compared with MISO DoA baseline in 3 and 4 sources scenes, and the experimental results also show MIMO-DoAnet alleviates the threshold setting problem and solves the angle assumption problem effectively. In the future, we will try to detect whether there is a sound source in each output by analyzing the SPS distribution, so that MIMO could get rid of the threshold and further improve the performance.

\section{Acknowledgements}
This work was supported in part by the National Natural Science Foundation of China under Grant 62176182.
 
\bibliographystyle{IEEEtran}

\bibliography{mybib}


\end{document}